\documentclass[aps,preprint,superscriptaddress,nofootinbib]{revtex4}

\usepackage{graphicx}
\usepackage{epic}
\usepackage{eepic}
\usepackage{latexsym}

\newcommand{\eq}[1]{(\ref{#1})}
\newcommand{\be}{\begin{equation}}
\newcommand{\ee}{\end{equation}}
\newcommand{\bea}{\begin{eqnarray}}
\newcommand{\eea}{\end{eqnarray}}

\newcommand{\vs}[1]{\vspace{#1 mm}}
\newcommand{\hs}[1]{\hspace{#1 mm}}

\def\a{\alpha}
\def\b{\beta}
\def\cc{\gamma}
\def\C{\Gamma}
\def\d{\delta}

\def\e{\epsilon}

\def\fr{\frac}

\def\l{\lambda}

\def\m{\mu}
\def\n{\nu}

\def\s{\sigma}
\def\S{\Sigma}

\def\O{\Omega}

\def\del{\partial}

\let\bm=\bibitem
\def\nn{\nonumber}

\newcommand{\dA}{\dot{A}}
\newcommand{\dB}{\dot{B}}
\newcommand{\ddA}{\ddot{A}}
\newcommand{\ddB}{\ddot{B}}
\newcommand{\rt}{\tilde{r}}
\newcommand{\Xii}{{\cal L}}

\begin{document}

\title{Supergravity Solutions for Harmonic, Static and Flux S-Branes}

\author{Ali Kaya}
\email[]{ali.kaya@boun.edu.tr}
\affiliation{Bo\~{g}azi\c{c}i University, Department of
Physics, \\ 34342, Bebek, \.Istanbul, Turkey\vs{1.5}}
\affiliation{Feza G\"{u}rsey Institute, \\34684, \c{C}engelk\"{o}y
\.Istanbul, Turkey\vs{2}}

\date{\today}

\begin{abstract}

We seek S-brane solutions in $D=11$ supergravity which can be characterized
by a harmonic function $H$ on the flat transverse space. It turns out
that the Einstein's equations force $H$ to be a linear function of the
transverse coordinates. The codimension one $H=0$ hyperplane can be
spacelike, timelike or null and the spacelike case reduces to the
previously obtained SM2 or SM5 brane solutions. We then consider
static S-brane configurations having smeared timelike directions
where the transverse Lorentzian symmetry group is broken down to
its maximal orthogonal subgroup. Assuming that the metric functions depend on
a radial spatial coordinate, we construct explicit solutions in
$D=11$ supergravity which are non-supersymmetric and asymptotically flat.
Finally, we obtain spacelike fluxbrane backgrounds which have timelike
electric or magnetic fluxlines extending from past to future infinity.

\end{abstract}

\maketitle

\section{Introduction}

Spacelike S$p$-branes in string theory are topological defects which
exist only for a moment in time. In perturbation theory they arise
when the time coordinate obeys a Dirichlet boundary
condition and in world sheet conformal field theory (CFT) they
can be described by boundary states implementing the boundary
conditions \cite{s1}. S-branes can also be considered as time
dependent tachyonic  kink solutions of unstable D-brane world-volume
theories and they are expected to play the role of D-branes in
realizing $dS/CFT$ duality \cite{strominger}.

If S-branes are stable objects then one expects to have a
corresponding supergravity description for them. Indeed, in
\cite{s1,s2,s3} various time dependent supergravity solutions were
constructed for S-branes in different dimensions. It is remarkable
that the standard intersection rules for usual $p$-branes also arise
for S-branes \cite{alisadik} and one can further find solutions for
nonstandard intersections \cite{sadik} or intersections of
$p$-branes with S-branes \cite{int}. By analytical continuations of
black holes, completely regular S-brane solutions can also be
obtained \cite{reg1,reg2,reg3} (for other works on S-brane solutions
see, e.g., \cite{oth0,oth1,oth2,oth3,oth4,oth5,oth6,oth7,oth77,oth8}).

S$p$-branes in $D$-dimensions have transverse $SO(1,D-p-2)$
R-symmetry. The Lorentz invariant combination of the transverse
coordinates gives a null direction, and as shown in \cite{s3}, one
can construct solutions corresponding to interior and exterior
regions of the light cone which can be foliated by hyperbolic and de
Sitter spaces, respectively. On the other hand, for some physical
applications it would be interesting to consider solutions where the
Lorentzian symmetry is broken (in terms of Euclidean gauge theory
living on the S-brane that would correspond to breaking of
R-symmetry by giving vacuum expectation values to scalars). One aim
of this paper is to study this possibility.

It is well known that usual $p$-brane solutions are characterized by
harmonic functions. In general the corresponding antisymmetric
tensor $F$ obeys $dF=0$ and $d*F=0$, and these can be satisfied
naturally using harmonic functions by writing $F\sim dH$ or $*F\sim dH$
for electric or magnetic solutions, respectively.
As long as the form fields are concerned S-branes are not very
different than $p$-branes. On the other hand, harmonic superposition
of $p$-branes is possible due to supersymmetry which is absent for
S-branes. In the next section we seek S-brane solutions which are
characterized by harmonic functions. If harmonic S-branes exist then
there should be a way of avoiding superposition principle. Indeed,
as we will see below, the form equations can be satisfied naturally
by harmonic functions but the Einstein's equations demand linearity
on the transverse space so that superposition of different
functions does not yield a new solution but simply modifies the parameters
in the total function. In the same section we also generalize harmonic
S-branes where the flat transverse spaces are replaced by
general Ricci flat manifolds.

For usual $p$-brane backgrounds, one common procedure is to smear a
transverse coordinate by putting an array of parallel branes in that
direction. In principle supersymmetry is required for smearing and
it is not clear how this may be achieved for S-branes. In any case,
one can still imagine a solution for S-branes distributed
uniformly in time (one can argue that stability
is not a question here since "time evolution" is fixed by hand
from the beginning like imposing a boundary condition).
For such a background there is a time translation symmetry and the
corresponding solution should be static. In section \ref{III}, we
construct explicit solutions in $D=11$ supergravity which can be thought to
represent time smeared S-branes where
the transverse Lorentzian R-symmetry group is  broken down to its
maximal  orthogonal subgroup. As we will see below these backgrounds
are non-supersymmetric, asymptotically flat, generically singular in
the interior but support finite ADM masses per unit Euclidean S-brane
volumes.

Finally in section \ref{IV} we obtain spacelike fluxbrane solutions in
$D=11$ supergravity theory. Generically, a fluxbrane background has
antisymmetric tensor field components tangent to the transverse
coordinates. Although the fluxlines have infinite extend the total
charge is finite. This is interpreted as the confinement of
the fluxlines by their own gravitational field.
The classical example is the Melvin solution
of 4-dimensional Einstein-Maxwell gravity \cite{mel} which describes a
flux 1-brane. Later various higher dimensional generalizations of fluxbranes
were constructed in the literature (see e.g. \cite{f1,f2,f3,f4,f5,f6,f7,f8}).
As the spacelike branes have recently attracted some attention in string theory,
one may wonder if there are solutions for spacelike fluxbranes.
In such a background the transverse space is Lorentzian and therefore
the antisymmetric tensor field should have a component along time direction.
As we will see one can construct time dependent solutions with this property,
representing spacelike fluxbranes. Similar to usual timelike fluxbranes,
the spacelike solutions have fluxlines extending from past to future
timelike infinity but the total flux still converges despite the infinite range.
This shows that the confinement of fluxlines by gravity also
works for spacelike solutions.

\section{Harmonic S-branes \label{II}}

In this section our aim is to construct S-brane solutions characterized by
harmonic functions. As an example let us consider SM2 brane in eleven
dimensions. The equations of motion for the bosonic fields of $D=11$
supergravity can be written as
\bea
&&R_{MN}=\fr{1}{3}F_{MPQR}F_{N}{}^{PQR}
-\fr{1}{36}g_{MN}\,F_{PQRS}\,F^{PQRS} \nn\\
&&dF=0,\hs{5}d*F=F\wedge F.\label{11e}
\eea
We consider the following metric\footnote{As shown in \cite{ak}
 as long as the supergravity solutions are concerned the "internal"
flat spaces, spheres or hyperboloids
can be replaced by arbitrary Ricci flat, positively or negatively curved
Einstein spaces, respectively. Therefore in \eq{2}, for instance,
the flat space spanned by $(x^1,x^2,x^3)$ can be replaced by any Ricci flat manifold.}
\be\label{2}
ds^2=e^{2A}\,\d_{ab}dx^adx^b\,+\,e^{2B}\,\eta_{\m\n}\,dy^\m dy^\n,
\ee
where $a,b=1,2,3$ and $\m,\n=0,..,7$. The coordinate
dependencies of the metric functions $A$ and $B$ are assumed to be
of the form $A=A(H)$ and $B=B(H)$, where $H=H(y)$ is a function on
the transverse space. For the antisymmetric tensor we take
\be\label{3}
F_{abc\m}=\fr{1}{2}\,e^{-7B}\,\e_{abc}\,\del_\m H.
\ee
Note that $dF=0$ is identically satisfied and the antisymmetric tensor equation
$d*F=0$ gives
\be\label{har}
\del_\m\,\del^\m \,H=0.
\ee
Using \eq{har} in Einstein's equations we find
\bea
&&\ddot{A}+3 \dot{A}^2+6 \dA \dB +\fr{e^{-12B}}{3}=0, \label{e1}\\
&&\ddB +6\dB^2+3\dA\dB-\fr{e^{-12B}}{6}=0, \label{e2}\\
&&\ddA+2\ddB+\dA^2-2\dB^2-2\dA\dB+\fr{e^{-12B}}{6}=0,\label{e3}\\
&&\dA+2\dB=0,\label{e4}
\eea
where dot denotes differentiation with respect to the
argument (i.e. $H$). Eq. \eq{e1} follows from the worldvolume directions
$(M,N)=(a,b)$ in \eq{11e} and the
terms coming from $(M,N)=(\m,\n)$ components fall into three different groups
with coefficients  $\eta_{\m\n}(\del H)^2$, $\del_\m H\del_\n H$ and
$\del_\m\del_\n H$. Setting them to zero separately\footnote{At this
  point we do not want to impose any additional condition on $H$,
  thus the functions $\eta_{\m\n}(\del H)^2$, $\del_\m H\del_\n H$ and
  $\del_\m\del_\n H$ are assumed to be linearly independent.} give
\eq{e2}, \eq{e3} and \eq{e4}, respectively. Eq. \eq{e4} yields $A=-2B$
and then \eq{e3} implies
\be\label{bos}
\dB^2+\fr{e^{-12B}}{36}=0,
\ee
which shows that the solution space is {\it empty}. This result is not
surprising since otherwise harmonic superposition of S-branes would be possible
which is odd in the absence of supersymmetry.

One way of proceeding is to consider type IIA* or IIB* string theories
studied in \cite{hull1,hull2} which can be obtained by timelike
T-dualities from type IIA and IIB strings, respectively.
In supergravity actions a timelike T-duality alters the signs of the
kinetic terms of the antisymmetric
tensor fields. Lifting IIA* theory to eleven
dimensions and considering the same ansatz \eq{2} and \eq{3}
in this framework, \eq{bos} becomes
\be
\dB^2-\fr{e^{-12B}}{36}=0,
\ee
which implies $B=(\ln H)/6$. Using also $A=-2B$, one ends up
with the Wick rotated Euclidean $p$-branes studied in \cite{hull1}.

Another possibility is to impose $\del_\m\del_\n H=0$ so that
\eq{e4} is dropped from the equation system. In this case the harmonic function
$H$ becomes
\be\label{harl}
H=c_\m y^\m + c,
\ee
where $(c_\m,c)$ are
constants. A systematic way of solving equations like
\eq{e1}-\eq{e3} was discussed in \cite{gen1}. Adding \eq{e1} and two times
\eq{e2} yields
\be
(\ddot{A}+2\ddot{B})+3(\dA+2\dB)^2=0,
\ee
which can be solved as $A+2B=(\ln H)/3$. Using this in \eq{e2} one
obtains
\be
\ddB+\fr{\dB}{H}-\fr{e^{-12B}}{6}=0,
\ee
which has the solution
\be
B=\fr{1}{12}\ln\left[b\, H^2\,
\cosh^2 \left[\fr{\ln (\pm H)}{\sqrt{b}}\right]\right],
\ee
where $b$ is a constant. Eq. \eq{e3}, on the other hand, acts
like a constraint equation which fixes $b=3/7$.
The end result is that
\bea
&&A=-\fr{1}{6}\ln\left[\fr{3}{7}
 \cosh^2 \left[\sqrt{\fr{7}{3}}\ln (\pm H)\right]\right],\nn\\
&&B=\fr{1}{12}\ln\left[\fr{3H^2}{7}
\cosh^2 \left[\sqrt{\fr{7}{3}}\ln (\pm H)\right]\right], \label{sm2}
\eea
where $\pm$ signs are for $H>0$ and $H<0$ regions respectively.

Let $\S_H$ be the codimension one $H=0$ hyperplane which
can be spacelike, timelike or
null. In each case one can apply suitable Lorentz transformations
and translations to set $H=c_0 t$, $H=c_1 y^1$ or $H=c_{\pm}(t\pm y^1)$,
respectively. After a coordinate transformation $\tilde{t}=\ln (c_0
t)$, it is not difficult to see that the spacelike case is
identical to the time dependent SM2 brane solution with a flat
transverse space. Here one discovers two additional
backgrounds corresponding to timelike and null planes. Note that
superposition of different harmonic functions (i.e. when
$H=\sum_i H_i$) does not give a new solution but simply
modifies the constants $(c_\m,c)$. This is somehow expected in
the absence of supersymmetry.

For the null plane, \eq{e2} is redundant since it is actually multiplied by
$\eta_{\m\n}\,(\del H)^2=0$ and the solution space is larger.
However, superposition with the spacelike and the timelike
backgrounds is not possible for this general class. So we impose \eq{e2}
as an additional equation to ensure superposition.

The electric charge of SM2 brane is given by
\be
Q=\int *F=\int_{\S_H} \hat{*}\,dH,
\ee
where $\hat{*}$ is the 8-dimensional Hodge dual on the flat transverse
space. The integrand is a constant and the total charge diverges. The
fluxlines are confined on $\S_H$, but they have constant
magnitude and infinite extend (of course it is possible to compactify
$\S_H$ to get finite charge). The metric is singular on $\S_H$
and at infinity along the perpendicular
direction to $\S_H$, i.e. when $H\to \pm\infty$.

It is easy to repeat the above construction for the magnetic SM5 brane
which has the following metric
\be
ds^2=e^{2A(H)}\,\d_{ab}dx^adx^b\,+\,e^{2B(H)}\,\eta_{\m\n}\,dy^\m dy^\n,
\ee
where $a,b=1,..,6$ and $\m,\n=0,..,4$. The antisymmetric tensor can be taken as
\be
F=\fr{1}{2}\,\hat{*}\, dH,
\ee
where $\hat{*}$ is the Hodge dual on the flat 5-dimensional transverse space.
The form equation $d*F=0$ is identically satisfied and $dF=0$ implies
$\del_\m\del^\m H=0$. On the other hand the Einstein's equations
require $\del_\m\del_\n H=0$ and thus $H$ is linear as in \eq{harl}.
The metric functions can be solved as
\bea
&&A=-\fr{1}{12}\ln\left[\fr{3}{8}
\cosh^2 \left[\sqrt{\fr{8}{3}}\ln (\pm H)\right]\right],\nn\\
&&B=\fr{1}{6}\ln\left[\fr{3\,H^2}{8}
\cosh^2\left[ \sqrt{\fr{8}{3}}\ln (\pm H)\right]\right]. \label{sm5}
\eea
Again $\pm$ signs are for $H>0$ and $H<0$ regions respectively.
The solution carries a magnetic charge
\be
Q=\int F=\int_{\S_H}\,\hat{*}\, dH,
\ee
which diverges since the constant magnetic fluxlines extend on $\S_H$
to infinity. When $\S_H$ is spacelike one can apply a Lorentz
transformation to set $H=c_0 t$ and the solution becomes the usual
time dependent SM5 brane with a flat transverse space after the
coordinate change  $\tilde{t}=\ln (c_0 t)$.

For both SM2 and SM5 branes R-symmetry groups are
determined by the isometries of the corresponding plane
$\S_H$. For instance for SM2 brane the symmetry groups are $ISO(7)$,
$ISO(1,6)$ and ${\cal R}\times ISO(6)$ when $\S_H$ is spacelike,
timelike and null, respectively.

It is known that time dependent solutions (i.e. when $\S_H$ is
spacelike in our case)  cannot be supersymmetric since the
"Hamiltonian", which can be written as the anti-commutator of
supercharges, is not a "constant". To see if there is any unbroken
supersymmetry for other cases let us consider the Killing spinor equation
\be\label{cd} D_M\e\equiv
\left[\nabla_M+\fr{1}{144}\left(\C^{PQRS}{}_M
-8\,\d^P_M\,\C^{QRS}\right)F_{PQRS}\right]\,\e =0.
\ee
We check out the integrability condition
\be\label{int}
 D_{[M}D_{N]}\e=0
\ee
and find that there is no Killing spinor when $\S_H$ is timelike or
spacelike. For instance in the SM2 brane solution $D_{[a}D_{b]}\e=0$
implies
\be\label{in1}
\left[\dA^2-\fr{e^{-12B}}{9}\right]\, \left(\del^\m H\del_\m
H\right)\,\C_{ab}\,\,\e=0.
\ee
Using \eq{sm2} one finds that
$(\dA^2-e^{-12B}/9)\not=0$ and thus $\e=0$
since $\del^\m H\del_\m H\not=0$.

For the null solution \eq{in1} is identically satisfied and
the integrability condition $D_{[a}D_{\n]}\e=0$
implies
\be\label{pr}
\C^{\m}\del_\m H\,\,\e=0.
\ee
Imposing \eq{pr}, one can find Killing spinors of the form
\be
\e=\begin{cases}{\left[\C^{\m}\del_\m H\right]\,e^{f+g\C^{123}}\,\e_0\hs{9}
\textrm{SM2},\cr\cr
\left[\C^{\m}\del_\m H\right]\,
e^{f+g\,\C^{123456}}\,\e_0\hs{5}
\textrm{SM5},}
\end{cases}
\ee
where  $\e_0$ is a constant spinor, $f=B/2$, $\dot{g}=e^{-6B}/12$
for SM2 and $\dot{g}=e^{-3B}/12$ for SM5 branes. Therefore the null
backgrounds preserve 16 supersymmetries of $D=11$ supergravity.

We should remark that the above solutions should be considered
on $H>0$ or $H<0$ regions separately. If one tries to use them
in the whole space then there appears extra delta function singularities
which arise since the metric functions \eq{sm2} and \eq{sm5} are discontinues
on $\S_H$. For both SM2 and SM5 branes the Ricci tensor components
containing second derivatives of the functions $A$ and $B$ can be found as
\bea
&&R_{ab}=-e^{-2B}\ddot{A}\,(\del_\l H)(\del^\l H)\d_{ab}+...\nn\\
&&R_{\m\n}=-e^{-2B}\ddot{B}\,(\del_\l H)(\del^\l H)\eta_{\m\n}+...
\eea
where the dotted terms involve only the first derivatives plus
the second derivatives of $(A+2B)$ for SM2 and $(2A+B)$ for SM5 branes
which are however continuous across $\S_H$.
From \eq{sm2} and \eq{sm5} one finds as $H\to0$ that
\be
e^{-2B}\ddot{A}\sim e^{-2B}\ddot{B} \sim
\begin{cases}{
H^{(-4+\sqrt{7/3})/3}\,\,\d(H)+...\hs{6}\textrm{SM2,}\cr\cr
H^{(-5+2\sqrt{8/3})/3}\,\,\d(H)+...\hs{5}\textrm{SM5.}}
\end{cases}
\ee
Therefore to obtain supergravity solutions in the whole region,
these backgrounds should be supplemented by additional delta function
sources on $\S_H$ which may arise from the coupling of elementary
S-branes to the supergravity fields. In this case, however, one
would expect the antisymmetric tensor fields also to be modified.
To achieve this one may replace \eq{harl} with $H=|c_\m y^\m+c|$,
which is the Greens function in one dimension perpendicular to $\S_H$.
Although there is no need for the negative signs in the metric functions
\eq{sm2} and \eq{sm5}, the second derivatives of $H$ now yield
additional delta functions in the Ricci tensor. Similarly, unless $c_\m $ is a null
vector the form equations are modified such that $d*F\sim\d(H)$ for SM2 and
$dF\sim\d(H)$ for SM5 branes.

As a final comment let us point out that it is possible to replace the flat
transverse space with a curved one
\be
\eta_{\m\n}\,dy^\m dy^\n \to d\,\Xii^2,
\ee
where $\Xii$ is an arbitrary Ricci flat Lorentzian space. It is not difficult to
verify that the SM2 and SM5 brane backgrounds \eq{sm2} and \eq{sm5}
satisfy field equations  provided $H$ obeys
\be
\hat{\nabla}_\m\hat{\nabla}_\n H=0,
\ee
where $\hat{\nabla}_\m$ is the covariant derivative on $\Xii$.
If such a function $H$ exists on $\Xii$ then it yields a
covariantly constant vector field $k_\m=\hat{\nabla}_\m
H$. Conversely, for any covariantly constant vector field $k_\m$ one
can find a {\it local} function $H$ such that $k_\m=\hat{\nabla}_\m H$.
Moreover when $k_\m$ is hypersurface orthogonal then $H$ is globally
well defined. Therefore, one can write down a solution for each
covariantly constant vector field on $\Xii$ where the metric functions
\eq{sm2} or \eq{sm5} depend on the (local) potential $H$ and
the antisymmetric tensor fields become $*\,F=\hat{*}\,k$ for SM2 and
$F=\hat{*}\,k$ for SM5 branes, where $\hat{*}$ is Hodge dual on $\Xii$
and $k$ is the one-form corresponding to $k_\m$. The electric or the
magnetic charge of this solution can be calculated as
\be
Q=\int  \hat{*}\,k
\ee
which gives a finite result when the cycle dual
to one-form $k$ is compact.

\section{Static S-branes \label{III}}

It is well known that a transverse direction to a $p$-brane worldvolume
can be smeared out by placing a continuum array of parallel branes in that
direction. This can be achieved due to supersymmetry
which ensures stability. It is not clear whether one can place
two parallel S-branes separated by a finite time interval and thus whether
smearing is possible. Assuming this can be done one can
consider an infinite array of S-branes. It can be claimed that the
stability of this system is not an issue since the time evolution is
dictated by hand like imposing a boundary condition. In this section
we construct explicit solutions in $D=11$ supergravity which can be thought to
represent smeared S-branes.

Let us consider SM2 brane first. It is clear that the supergravity solution
should be static. Moreover the transverse $SO(1,7)$ symmetry should be
broken down to SO(7) subgroup and one can introduce a radial
transverse coordinate. Thus the metric can be taken as
\be
ds^2=e^{2A}\left(dx_1^2+dx_2^2+dx_3^2\right)-e^{2B}dt^2+e^{2C}dr^2
+e^{2D}d\O_6^2,
\ee
where $\O_6$ is the unit 6-sphere and the unknown functions depend on
$r$. The three-form potential ${\cal A}$ should couple to the
Euclidean world volume and can be written as ${\cal A}_{abc}=f(r)\e_{abc}$.
Solving the antisymmetric tensor equation $d*F=0$  one finds
(the indices refer to the tangent space)
\be
F_{abcr}=\fr{k}{2}\,e^{-B-6D}\,\e_{abc},
\ee
where $k$ is an integration constant. Let us point out that unlike the similarity
the above background differs from brane anti-brane
systems studied in the literature   (see e.g. in
\cite{gen1,gen2,gen3}). The main distinction is in the choice of the
antisymmetric tensor which is constructed here to represent a
Euclidean brane.

Imposing the gauge
\be
C=3A+B+6D,
\ee
one finds the following second order equations
\bea
&&A''=-\fr{k^2}{3}\,e^{6A},\nn\\
&&B''=\fr{k^2}{6}\,e^{6A},\label{se}\\
&&D''=5\,e^{6A+2B+10D}\,+\,\fr{k^2}{6}\,e^{6A},\nn
\eea
together with a first order constraint
\be\label{cons}
C'^2-3A'^2-B'^2-6D'^2-\fr{k^2}{2}\,e^{6A}-30\,e^{6A+2B+10D}=0,
\ee
where prime denotes differentiation with respect to $r$. The system
\eq{se} can be integrated step by step starting from $A$ to yield
\bea
&&A=-\fr{1}{3}\ln\left[k\cosh r\right],\nn\\
&&B=\fr{1}{6}\ln\left[k\cosh r\right]+c_1r,\label{sol1}\\
&&D=\fr{1}{6}\ln\left[k\cosh r\right]-
\fr{1}{5}\ln\left[\fr{10}{c_2}\sinh\left(\fr{c_2\,r}{2}\right)\right]
-\fr{c_1r}{5},\nn
\eea
where $c_1$ and $c_2$ are integration constants (we scale $r$ to
eliminate one constant in $A$). The constraint \eq{cons} imposes
\be\label{c1c2}
c_2^2\,=\,\fr{5}{3}+4\,c_1^2.
\ee
One should choose $c_2>0$ to have a well defined metric function $D$ in \eq{sol1}
and $c_1$ can be positive, negative or
zero. Introducing a new radial coordinate
\be
\rt=\left[\tanh\left(\fr{c_2\,r}{4}\right)\right]^{-1/5},
\ee
the metric becomes
\bea\label{met2}
ds^2=e^{2A}\left(dx_1^2+dx_2^2+dx_3^2\right)-e^{2B}dt^2
+\fr{e^{2D}}{\rt^2}\left[d\rt^2+\rt^2d\O_6^2\right],
\eea
where the functions $A$, $B$ and $D$ are still given by \eq{sol1} with
$r=4\,\,\textrm{arctanh}(\rt^{-5})/c_2$. In \eq{met2} one can
introduce Cartesian coordinates in the flat space parametrized by
$(\rt,\O_6)$ and $S0(7)$ symmetry, which acts as rotations
around the fixed origin, becomes manifest.

The above solution is asymptotically flat as $\rt\to\infty$ (or $r\to 0$)
where the metric functions can be expanded as
\bea
&&e^A=k^{-1/3}
\left[1+{\cal O}(\fr{1}{\rt^{10}})\right]\nn\\
&&e^B=k^{1/6}
\left[1+\fr{4c_1}{c_2}\,\,\fr{1}{\rt^5}+{\cal O}(\fr{1}{\rt^{10}})\right]\\
&&e^D=k^{1/6}\left(\fr{c_2}{20}\right)^{1/5} \rt
\left[1-\fr{4c_1}{5c_2}\,\,\fr{1}{\rt^5}+{\cal O}(\fr{1}{\rt^{10}})\right].\nn
\eea
Here $1/\rt^5$ fall off is expected since the spatial transverse
space is 7-dimensional. The solution supports
finite ADM mass (per unit Euclidean volume) which is given by
\be
M=\O_6\,k^{5/6}\,\fr{3c_1}{5\kappa^2},
\ee
where $\kappa$ is the gravitational coupling constant and $\O_6$ is the
volume of unit 6-sphere. To get positive mass one should choose
$c_1>0$. Although $c_2$ does not contribute to ADM mass it
cannot be scaled away.

To analyze the interior region near $\rt\to1$ (or $r\to\infty$) let us
introduce a new coordinate $u$ with
\be
u= \fr{e^{-c\,r}}{c}, \hs{10}
c=\fr{3c_2+c_1}{5}-\fr{1}{6}.
\ee
We note that due to \eq{c1c2} $c$ is always a positive
constant. In the limit $\rt\to1$, (or $r\to\infty$), $u\to0$ and the metric becomes
\be
ds^2\to
u^{2/(3c)}\left(dx_1^2+dx_2^2+dx_3^2\right)-u^{-(1+6c_1)/(3c)}
dt^2+du^2+u^{2-c_2/c}d\O_6^2,
\ee
which is singular at $u=0$ since for instance the coefficient of
$dx_1^2$ vanishes. Although the solution is asymptotically
flat and supports finite ADM mass it contains a naked singularity in
the interior.

To see whether there is any unbroken supersymmetry in the system we
check out the integrability condition \eq{int}. A simple calculation shows that
$D_{[a}D_{t]}\e=0$ implies
\be
\C^{at}\,\left(f+g\,\C^{123}\,\right)\e=0 \hs{5}\Rightarrow \hs{5}\e=0,
\ee
where $f=-A'B'e^{A+B-2C}/4$ and $g=kB'e^{A-C-6D}/24$. Since
$\left(f+g\C^{123}\right)$ is an invertible matrix the solution does not
preserve any supersymmetry.

It is possible to smear some of the transverse directions and consider
a metric of the form\footnote{One can consider the most general case
  where each transverse $y$-coordinate in \eq{ssol} is multiplied by a different
  function. It turns out that the equation system can still be decoupled in this
general background.}
\be\label{ssol}
ds^2=e^{2A}\left(dx_1^2+dx_2^2+dx_3^2\right)-e^{2B}dt^2+e^{2C}dr^2
+e^{2D_1}\left(dy_1^2+..+dy_m^2\right)+e^{2D_2}d\O_n^2,
\ee
where $m+n=6$ and $n\geq 2$. The antisymmetric tensor should be modified as
\be
F_{abcr}=\fr{k}{2}\,e^{-B-mD_1-nD_2}\,\e_{abc}.
\ee
Imposing the gauge
\be
C=3A+B+mD_1+nD_2,
\ee
the differential equations are decoupled and the metric
functions can be integrated step by step to yield
\bea
&&A=-\fr{1}{3}\ln\left[k\cosh r\right],\nn\\
&&B=\fr{1}{6}\ln\left[k\cosh r\right]+c_1r,\nn\\
&&D_1=\fr{1}{6}\ln\left[k\cosh r\right]+c_2r,\label{sol2}\\
&&D_2=\fr{1}{6}\ln\left[k\cosh r\right]-
\fr{1}{(n-1)}\ln\left[\fr{2(n-1)}{c_3}\sinh\left(\fr{c_3\,r}{2}\right)\right]
-\fr{c_1+(6-n)c_2}{(n-1)}\,r,\nn
\eea
where the constants obey
\be
n\,c_3^2-20\,m\,c_2^2-4\,n\,c_1^2-8\,m\,c_1\,c_2+2-2n=0.
\ee
If one chooses $c_1=c_2$ then $B=D_1$, i.e. the metric
factors multiplying $t$ and $(y_1,...,y_m)$ coordinates become
equal. In this case there is an extra $ISO(1,m)$ symmetry acting on the
space spanned by $(t,y_1,..,y_m)$. The proper radial coordinate is given by
$\rt=\tanh(c_3r/4)^{-1/(n-1)}$ such that in \eq{ssol}
\be
e^{2C}dr^2+e^{2D_2}d\O_n^2\to\fr{e^{2D_2}}{\rt^2}\left(d\rt^2+\rt^2 d\O_n^2\right).
\ee
The solution is still asymptotically flat
as $\rt\to\infty$ (or $r\to0$) and singular in the interior as
$\rt\to1$(or $r\to\infty$). The metric functions fall off at least with the
power $\rt^{-(n-1)}$ so that the ADM mass is finite.

Let us now consider the static SM5 brane solution.
The metric and the antisymmetric tensor are given by
\bea
&&ds^2=e^{2A}\left(dx_1^2+..+dx_6^2\right)-e^{2B}dt^2+e^{2C}dr^2
+e^{2D}d\O_3^2,\nn\\
&&F_{\a\b\cc t}=\fr{k}{2}\,e^{-B-3D}\,\e_{\a\b\cc},
\eea
where the indices $\a,\b,\cc$ refer to the tangent space on $\O_3$ and the
unknown functions depend on $r$. The field equations can be solved to get
\bea
&&A=-\fr{1}{6}\ln\left[k\cosh r\right],\nn\\
&&B=\fr{1}{3}\ln\left[k\cosh r\right]+c_1\,r,\label{solm5}\\
&&C=\fr{1}{3}\ln\left[k\cosh r\right]-
\fr{3}{2}\ln\left[\fr{4}{c_2}\sinh\left(\fr{c_2\,r}{2}\right)\right]
-\fr{c_1r}{2},\nn\\
&&D=\fr{1}{3}\ln\left[k\cosh r\right]-
\fr{1}{2}\ln\left[\fr{4}{c_2}\sinh\left(\fr{c_2\,r}{2}\right)\right]
-\fr{c_1r}{2},\nn
\eea
where the constants obey
\be
c_2^2=\fr{4}{3}+4c_1^2.
\ee
Introducing the proper radial coordinate
\be
\rt=\left[\tanh\left(\fr{c_2\,r}{4}\right)\right]^{-1/2},
\ee
the metric becomes
\be
ds^2=e^{2A}\left(dx_1^2+..+dx_6^2\right)-e^{2B}dt^2+\fr{e^{2D}}{\rt^2}
\left(d\rt^2 +\rt^2\,d\O_3^2\right).
\ee
The $SO(4)$ R-symmetry acts on the flat space spanned by
$(\rt,\O_3)$. The solution is asymptotically flat as $\rt\to\infty$
(or $r\to 0$) where the functions $e^A$, $e^B$ and $\rt e^D$ fall off
with the powers $1/\rt^4$, $1/\rt^2$ and $1/\rt^2$, respectively. The
ADM mass (per unit Euclidean volume) can be calculated as
\be
M=\O_3\,k^{2/3}\,\fr{3c_1}{4\kappa^2},
\ee
where $\kappa$ is the gravitational coupling constant and $\O_3$ is the
volume of unit 3-sphere. The metric is singular as $\rt\to 1$ (or $r\to\infty$, $u\to0$):
\be
ds^2\to
u^{1/(3c)}\left(dx_1^2+dx_2^2+dx_3^2\right)-u^{-(2+6c_1)/(3c)}
dt^2+du^2+u^{2-c_2/c}d\O_6^2,
\ee
where $u=e^{-cr}/c$ and $c=-1/3+c_1/2+3c_2/4$ is a positive constant.
We check out that there is no Killing spinor on this
background and thus the solution is not supersymmetric.

Finally let us note the smeared solution where $\O_3\to {\cal R}\times\O_2$. The fields are
given by
\bea
&&ds^2=e^{2A}\left(dx_1^2+..+dx_6^2\right)-e^{2B}dt^2+e^{2C}dr^2
+e^{2D_1}dy^2+e^{2D_2}d\O_2^2,\nn\\
&&F_{\a\b y t}=\fr{k}{2}\,e^{-B-D_1-2D_2}\,\e_{\a\b\cc},
\eea
where
\bea
&&A=-\fr{1}{6}\ln\left[k\cosh r\right],\nn\\
&&B=\fr{1}{3}\ln\left[k\cosh r\right]+c_1\,r,\nn\\
&&C=\fr{1}{3}\ln\left[k\cosh r\right]-
2\ln\left[\fr{2}{c_3}\sinh\left(\fr{c_3\,r}{2}\right)\right]
-(c_1+c_2)r,\label{solmsmear5}\\
&&D_1=\fr{1}{3}\ln\left[k\cosh r\right]+c_2\,r,\nn\\
&&D_2=\fr{1}{3}\ln\left[k\cosh r\right]-
\ln\left[\fr{2}{c_3}\sinh\left(\fr{c_3\,r}{2}\right)\right]
-(c_1+c_2)r,\nn
\eea
and the constants $c_1,c_2,c_3$ obey
\be
c_3^2-4\,c_2^2-4\,c_1^2-4\,c_1\,c_2-1=0.
\ee
The proper radial coordinate is
$\rt=\tanh(c_3r/4)^{-1}$. The metric is asymptotically flat as
$\rt\to\infty$ and it is singular
in the interior as $\rt\to1$.

\section{Spacelike Fluxbranes\label{IV}}

The main example of a fluxbrane is the Melvin background of
4-dimensional Einstein-Maxwell gravity \cite{mel} which is given by
\bea
&&ds^2=\left(1+\fr{B^2r^2}{4}\right)^2\left[-dt^2+dz^2+dr^2\right]
+\fr{r^2}{\left(1+\fr{B^2r^2}{4}\right)^2}
\,\, d\phi^2\nn\\
&&F=\fr{B}{\left(1+\fr{B^2r^2}{4}\right)^2}\,\,r\,dr\wedge d\phi.
\eea
The constant $B$ is the magnetic field strength on the axis $r=0$. The
total flux can be calculated as
\be
\int_{R^2} F =\fr{4\pi}{B},
\ee
which is finite although the lines have infinite extend. This is
interpreted as the confinement of the magnetic fluxlines by
gravity. As $r$ increases the orbits of $\phi$ become small and the
solution resembles a teardrop with an infinite tail.

To get a spacelike fluxbrane in this theory we perform the following
analytical continuations
\be
r\to i\, t,\hs{5}t\to  i\,y,\hs{5} B\to i\, B,\hs{5} \phi\to i\,\phi,
\ee
which give
\bea
&&ds^2=\left(1+\fr{B^2t^2}{4}\right)^2
\left[-dt^2+dy^2+dz^2\right]+\fr{t^2}{\left(1+\fr{B^2t^2}{4}\right)^2}
\,\,d\phi^2\nn\\
&&F=\fr{B}{\left(1+\fr{B^2t^2}{4}\right)^2}\,\,t\,dt\wedge d\phi.
\eea
This can be interpreted as a Euclidean flux 1-brane which has the
worldvolume coordinates $(y,z)$.  The fluxlines now extend in time to
infinity but the integral of $F$ is still finite. The geometry is
locally flat as $t\to 0$ and the orbits of $\phi$ diminish
as $t\to\infty$, resembling a teardrop extending in time.

Our aim in this section is to construct higher dimensional
generalizations of spacelike fluxbranes. Let us consider flux SM3
brane first which has the metric
\be\label{sm3}
ds^2=e^{2A}\left(dx_1^2+..+dx_4^2\right)-e^{2B}dt^2+e^{2C}d\S_{6}^2,
\ee
where the functions $A,B,C$ depend on $t$ and $\S_6$ is a Ricci flat
(preferably compact) space. For the antisymmetric tensor we take
\be
F_{abcd}=\fr{k}{2}\,e^{-4A}\,\e_{abcd},
\ee
so that the form equations are identically satisfied. This is an
electrically charged solution and the fluxlines (related to $*F$)
both extend in time and wrap over $\S_6$.

Imposing the gauge $B=4A+6C$ the field equations become
\bea
&&\ddot{A}=\fr{k^2}{3}\,e^{12C}\nn\\
&&\ddot{C}=-\fr{k^2}{6}\,e^{12C}\label{sm3eq}\\
&&10\dot{A}^2+30\dot{C}^2+48\dot{A}\dot{C}-\fr{k^2}{2}e^{12C}=0,\nn
\eea
where dot denotes time derivative. These equations
can be integrated to get
\bea
&&A=\fr{1}{3}\ln\left[k\cosh t\right]\pm \fr{t}{2\sqrt{6}},\nn\\
&&B=\fr{1}{3}\ln\left[k\cosh t\right]\pm \fr{4t}{2\sqrt{6}},\\
&&C=-\fr{1}{6}\ln\left[k\cosh t\right].\nn
\eea
The constant $k$ is related to the field strength at $t=0$. As
$t\to\pm\infty$, $C\to -\infty$ and thus the transverse space
is like the tail of an infinite tear drop (that has the shape of
$\S_6$) appearing in time. The total electric flux is given by
\be
\int *F=\fr{k}{2}\, V_6 \,\int_{-\infty}^{+\infty}e^{-4A+B+6C}dt=\fr{V_6}{k},
\ee
where $V_6$ is the volume of $\S_6$. Therefore, when $\S_6$ is compact,
the total charge is finite.

In a similar way one can construct flux SM6 brane which is given
by
\bea
&&ds^2=e^{2A}\left(dx_1^2+..+dx_7^2\right)-e^{2B}dt^2+e^{2C}d\S_{3}^2,\nn\\
&&F_{\a\b\cc t}=\fr{k}{2}\,e^{-7A}\,\e_{\a\b\cc},
\eea
where the indices $\a,\b,\cc$ refer to the tangent space of the Ricci flat manifold
$\S_3$. The metric functions can be found as
\bea
&&A=\fr{1}{6}\ln\left[k\cosh t\right]\pm \fr{t}{2\sqrt{21}},\nn\\
&&B=\fr{1}{6}\ln\left[k\cosh t\right]\pm \fr{7t}{2\sqrt{21}}, \\
&&C=-\fr{1}{3}\ln\left[k\cosh t\right],\nn
\eea
and the total magnetic charge is
\be
\int F=\fr{k}{2}\, V_3 \,\int_{-\infty}^{+\infty}e^{-7A+B+3C}dt=\fr{V_3}{k},
\ee
where $V_3$ is the volume of $\S_3$.

As discussed in \cite{f6}, there is an interplay between fluxbranes
and $p$-branes. Namely, a fluxbrane can be described as a limit of a
brane anti-brane system which is similar as the appearance of constant
electric field lines in between the plates of a capacitor. One would
expect a similar relation to hold for S-branes and flux
S-branes. Namely a flux S-brane should be realized as the limit of two
S-branes separated by a finite time interval. It would be interesting
to search for this possibility using the regular S-brane solutions
constructed in \cite{reg1,reg2,reg3}.

One can work out generalizations of the above solutions where the
Ricci flat space $\S$ is replaced by a positively or negatively
curved Einstein manifold (especially a solution with a sphere looks
natural for the compactness of fluxlines wrapping over it).
However, as in the case of usual fluxbrane backgrounds,
the differential equations cannot be decoupled due to the extra curvature terms.
For instance in the flux SM3 brane solution the equation system \eq{sm3eq} is modified such that
\bea
&&\ddot{A}=\fr{k^2}{3}\,e^{12C}\nn\\
&&\ddot{C}=-\fr{k^2}{6}\,e^{12C}-5\s e^{8A+10C}\\
&&10\dot{A}^2+30\dot{C}^2+48\dot{A}\dot{C}-\fr{k^2}{2}e^{12C}+30\s e^{8A+10C}=0,\nn
\eea
where $\s=+1$ and $\s=-1$ correspond to $\S$ being a unit sphere and a
unit hyperboloid in \eq{sm3}, respectively. It seems impossible to decouple
these equations due to $\s$ terms which are related to curvature of $\S$. In this case
one can try to integrate equations numerically or search for special exact solutions.
For $\s=-1$, we found a power law solution which can be written as
\bea
&&A=-\fr{1}{24}\ln \left(\a t \right)+\b,\nn\\
&&B=-\fr{7}{6}\ln \left(\a t\right)+4\b,\\
&&C=-\fr{1}{6}\ln \left(\a t\right),\nn
\eea
where $\a=2\sqrt{2k}$ and $e^{8\b}=3k^2/10$. Introducing the proper time coordinate
$d\tau=-e^{B}dt$ the metric becomes
\be\label{75}
ds^2=(\tilde{\a}\tau)^{1/2}\,e^{2\b}\,\left(dx_1^2+..+dx_4^2\right)-d\tau^2+
(\tilde{\a}\tau)^2\,dH_6^2,
\ee
where $\tilde{\a}=2(5/3)^{1/2}/3$. Eq. \eq{75} can be viewed as the asymptotic limit of
a more general solution and one can see that the integral
of $F$ converges as $\tau\to\infty$. Let us note that
the transverse space parameterized by ($\tau$,$H_6$) is {\it not} flat and there is a conic
singularity as $\tau\to0$ since $\tilde{\a}\not =1$. For $\s=+1$, the power law ansatz does not work
and we cannot find a special solution. It seems that numerical
techniques should be used to integrate equations for this case.

\section{Conclusions}

In recent discoveries on nonperturbative aspects of string theory, $p$-brane solutions
played a crucial role. Especially backgrounds corresponding to D-branes gave a lot of
new information since their dual CFT description as open strings obeying
Dirichlet boundary conditions are known. S-brane solutions are also expected to shed some light on
time dependent phenomena in string theory. In worldsheet CFT, S-branes arise when
the time coordinate obeys a Dirichlet boundary condition. In terms of supergravity fields
they can be described as time dependent solutions. Despite recent interesting developments
it seems that to have a better understanding of S-branes
more information is needed in both side of these dual descriptions.

In this paper, we construct new S-brane solutions in $D=11$ supergravity theory.
Firstly, we seek for solutions that can be characterized by a harmonic function $H$
on the transverse space. It turns out that the Einstein's equations demand $H$ to be a linear
function. The solutions can be classified
according to the codimension one hyperplane $\S_H$ being spacelike, timelike
or null. We observe that spacelike backgrounds are identical to the previously
constructed SM2 and SM5 brane solutions with flat transverse spaces.
Our construction reveals two additional family corresponding to timelike and null planes.
It is possible to superpose different solutions which would simply rotate or shift
the plane $\S_H$. The null solution preserves 16 supersymmetries of $D=11$ supergravity and
others are non-supersymmetric. We also show that the solutions can be generalized
naturally with arbitrary Ricci flat Lorentzian spaces.
The harmonic S-brane solutions can be thought as the generalizations of S-branes
with a flat transverse space. It would be interesting to consider other cases
with spherical or hyperbolic transverse spaces and find their harmonic extensions.

In supergravity brane solutions it is crucial to identify symmetries properly.
For an S$p$-brane in $D$ dimensions one would expect an $SO(1,D-p-2)$ symmetry
in the transverse space. However in some physical applications the symmetry groups
are necessarily broken down. In terms of gauge theories living on the branes
that would correspond to giving vacuum expectation values to some scalars.
In supergravity description symmetries are broken in the solutions when
parallel branes are separated from each other. It is not clear whether one can place two parallel
S-branes separated by a finite time interval. Assuming this can be done, one can smear
the time coordinate and $SO(1,D-p-2)$ symmetry should be broken down to $SO(D-p-2)$
subgroup. In this work we also construct solutions which can be thought
to represent time smeared static S-branes. These backgrounds resemble black $p$-brane
solutions in that they are asymptotically flat and support finite ADM masses. However
static S-branes are not black objects since they contain generic naked singularities
in the interior. They are also non-supersymmetric.

Finally, we obtain solutions for spacelike fluxbranes. A fluxbrane background
has antisymmetric tensor field components tangent to the transverse space.
The main characteristic property is the convergence of the total charge although
fluxlines have infinite extend. For a spacelike fluxbrane the transverse space is Lorentzian.
Not surprisingly the fluxlines now extend in time from past to future infinity
but the total charge is still finite. As for timelike fluxbranes one would
expect to obtain spacelike backgrounds as the limit of a solution which describes
S-brane pairs separated by a finite time interval.

Various S-brane solutions have been constructed in the literature and in this paper
we obtain new solutions which have interesting physical properties.
We believe however that the final word on supergravity description of S-branes is not
said. Especially, one should have a more clearer understanding of
the relation between supergravity solutions and CFT description of S-branes.
For an object that appears for a moment in time one would expect the corresponding
solution to be localized both in time and in transverse spatial coordinates.
In this case, however, it seems $SO(1,D-p-2)$  symmetry should be broken. It would be
interesting to study this possibility and construct purely localized S-brane
solutions.

\acknowledgments{I would like to thank Ata Karak\c{c}\i\ for his collaboration
in the beginning of this project and useful discussions.
This work is partially supported by Turkish Academy
of Sciences via Young Investigator Award Program (T\"{U}BA-GEB\.{I}P).}

\end{document}